\documentclass[12pt]{emulateapj}
\newcommand{\MJ}{M_{\rm J}}
\newcommand{\RJ}{R_{\rm J}}
\newcommand{\Qp}{Q_{\rm p}}
\newcommand{\Mp}{M_{\rm p}}
\newcommand{\Rp}{R_{\rm p}}
\newcommand{\Fp}{F_{\rm p}}

\newcommand{\Mc}{M_{\rm c}}
\newcommand{\ac}{a_{\rm c}}
\newcommand{\ec}{e_{\rm c}}

\begin{document}
\title{On The Origins Of Eccentric Close-in Planets}
%
\author{Soko Matsumura, Genya Takeda,}
\author{Frederic A.~Rasio}\affil{Department of Physics \& Astronomy,
Northwestern University, Evanston, IL, 60208}
%
%
\begin{abstract}
Strong tidal interaction with the central star can circularize the
orbits of close-in planets. With the standard tidal quality factor
$Q$ of our solar system, estimated circularization times for
close-in extrasolar planets are typically shorter than the ages of
the host stars. While most extrasolar planets with orbital radii
$a\la0.1\,$AU indeed have circular orbits, some close-in planets
with substantial orbital eccentricities have recently been
discovered. This new class of eccentric close-in planets  implies
that either their tidal $Q$ factor is considerably higher, or
circularization is prevented by an external perturbation.  Here we
constrain the tidal $Q$ factor for transiting extrasolar planets by
comparing their circularization times with accurately determined
stellar ages. Using estimated secular perturbation timescales, we
also provide constraints on the properties of hypothetical second
planets exterior to the known ones.
\end{abstract}
\section{Introduction}
\begin{figure}
\plotone{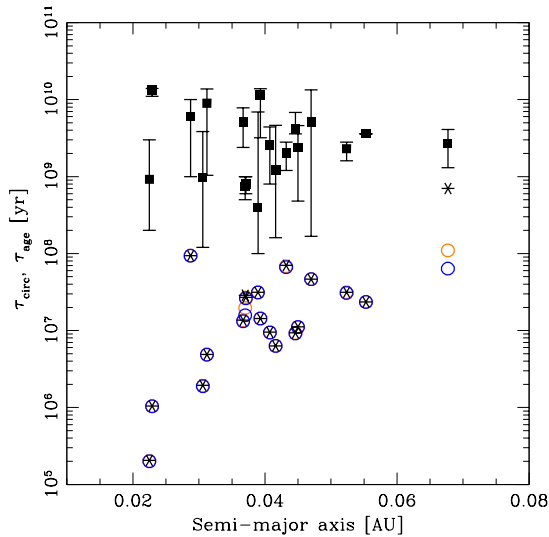} \caption[fig1]{ Circularization times calculated
from Eq.~\ref{tcirc0} (stars), Eq.~\ref{tcirc} with $\Omega_{*,{\rm
rot}}/n=3\,$d (orange circles) and 70\,d (blue circles),
compared with the estimated stellar ages (squares) for systems with
transiting planets. Here we assume $\Qp^{\prime}=10^5$ and
$Q_*^{\prime}=10^6$. For all transiting planets, the estimated
circularization time is shorter than the age of the host star.
\label{fig1}}
\end{figure}

The median eccentricity of the current sample of $\sim300$ planets
is 0.19, while it is 0.013 for close-in planets with semi-major axis
$a<0.1\,$AU. The circular orbits of close-in planets most likely
result from orbital circularization due to tides
\citep[e.g.,][]{Rasio96,Marcy97}. This requires the tidal
circularization time $\tau_{\rm circ}$ to be short compared to the
age of the system $\tau_{\rm age}$. Since $\tau_{\rm circ}$ is a
very steep function of $a$ (see Eq.~\ref{tcirc} or \ref{tcirc0}),
while $\tau_{\rm age} \sim 1-10\,$Gyr for most systems, a sharp
decline in eccentricity is expected below some critical value of
$a$. However, the observed transition seems to occur around
$0.03-0.04\,$AU, whereas the calculated \(\tau_{\rm circ}\) becomes
comparable to \(\tau_{\rm age}\) at $\sim 0.1\,$AU as we see below
(Figure~\ref{fig1}). Since almost one quarter (presently 16/68) of
planets within $0.1\,$AU have $e>0.1$, their high eccentricities
demand explanation.

First, we calculate the circularization times for transiting
planets, and compare them with the estimated ages of the systems.
The circularization time $\tau_{\rm circ}=-e/\dot{e}$, where
$\dot{e} $ is the sum of the eccentricity change due to the tides
raised on the star by the planet and those raised on the
planet by the star, is
\citep{Goldreich66,Hut81,Eggleton98,Mardling02}:
\begin{equation}
\tau_{\rm circ}=
\frac{2}{81}\frac{\Qp^{\prime}}{n}\frac{\Mp}{M_*}
\left(\frac{a}{\Rp}\right)^{5}
\left[\frac{\Qp^{\prime}}{Q_*^{\prime}}\left(\frac{\Mp}{M_*}\right)^2
\left(\frac{R_*}{\Rp}\right)^{5} F_* + \Fp \right]^{-1}
\label{tcirc} \,
\end{equation}
The subscripts p and $*$ represent the planet and star,
respectively. The modified tidal quality factor for a planet is
defined as $\Qp^{\prime}\equiv3\Qp/2k_{\rm p}$, where $k_{\rm p}$ is
the Love number, and $\Qp$ is the specific dissipation function, which
depends on the planetary structure as well as the frequency and
amplitude of tides. We also define
\begin{eqnarray}
F_*&=&\left[f_1(e^2)-\frac{11}{18}f_2(e^2)\frac{\Omega_{*,{\rm
rot}}} {n}\right]
\\
\Fp&=&\left[f_1(e^2)-\frac{11}{18}f_2(e^2)\frac{\Omega_{{\rm
p,rot}}} {n}\right] \ ,
\end{eqnarray}
where $n=\sqrt{G(M_*+\Mp)/a^3}$ is the mean motion, $\Omega_{\rm
rot}$ is the rotational frequency, and
\begin{eqnarray}
f_1(e^2)&=&
\left(1+\frac{15}{4}e^2+\frac{15}{8}e^4+\frac{5}{64}e^6\right)/(1-e^2)
^{13/2},
\\
f_2(e^2)&=&\left(1+\frac{3}{2}e^2+\frac{1}{8}e^4\right)/(1-e^2)^5 \
.
\end{eqnarray}
Generally $F_*$ and $F_p$ are comparable, and thus the stellar
damping is negligible unless the planet-to-star mass (radius) ratio
is large (small) or $Q_*^{\prime}\ll Q_p^{\prime}$. We define the
circularization time due to damping in the planet as
\begin{equation}
\tau_{{\rm
circ},0}=\frac{2}{81}\frac{\Qp^{\prime}}{n}\frac{\Mp}{M_*}
\left(\frac{\Rp}{a}\right)^{-5} \Fp^{-1} \  \label{tcirc0}.
\end{equation}
Note that $\tau_{{\rm circ},0}$ can be shorter or longer than
$\tau_{{\rm circ}}$, depending on the sign of $F_*$, which changes
at $(\Omega_{*,{\rm rot}}/n)_{\rm crit}= 18/11(f_1/f_2)$. In the
limit $e\rightarrow 0$, this equation leads to the standard
expression for the circularization time \citep[Eq.~4.198
of][]{Murray99}.

Figure~\ref{fig1} compares the circularization times calculated from
Eq.~\ref{tcirc} and \ref{tcirc0} with the estimated stellar ages for
the systems in Table~\ref{tb1}
\footnote{Note that, in Eq.~\ref{tcirc}, it is implicitly assumed
that the star and the planet both have zero obliquity. Currently
available measurements of the Rossiter--MacLaughlin effect show
that the planetary orbits in general are closely aligned with the
stellar equator \citep{Queloz00,Winn05}.  Current exceptions may be
the HD\,17156 and XO-3 systems \citep{Narita07ap,Hebrard08ap}, which
we exclude from our analysis.}.
%
Here we assume $\Qp^{\prime}=10^5$, and $Q_*^{\prime}=10^6$, which
are the standard values motivated by measurements in our Solar
System \citep[e.g.][]{Yoder81,Zhang08}, and for main-sequence stars
\citep[e.g.][]{Carone07}. For $\Omega_{\rm p,rot}$, we assume that
planets with circular orbits are perfectly synchronized, i.e.,
$\Omega_{\rm p,rot}/n=1$, since the spin-orbit synchronization times
are $\sim10^{-3}\,\tau_{\rm circ}$ \citep {Rasio96}. On the other
hand, planets with eccentric orbits should spin down until they
reach quasi-synchronization \citep{DobbsDixon04}; in practice we
adopt a planetary spin frequency such that the rate of change of
spin frequency is zero \citep[Eq.~54 of][]{Mardling02}. For the
stellar spin, we assume typical periods derived from the observed
$v_{\rm rot}\sin i$, $P_{*,{\rm rot}}\sim 3\,$--\,70\,d
\citep{Barnes01}.

Figure~\ref{fig1} compares the circularization times calculated from
Eq.~\ref{tcirc} and \ref{tcirc0} with the estimated stellar ages for
the systems in Table~\ref{tb1}.  For most systems the tidal damping
in the star is negligible. Two systems with non-negligible stellar
damping are WASP-14 and HAT-P-2. Both have a large planetary mass
(see Table~\ref{tb1}), and thus the first term in Eq.~\ref{tcirc} is
significant. Clearly, the estimated circularization times are always
shorter than stellar ages, which implies that all these planets
should have been circularized by now if their tidal $Q$ values were
similar to those of their solar-system analogues. However, our
sample contains at least 6 systems with a non-zero orbital
eccentricity. Possible explanations are that (1) the structure of
these planets is different and their actual $Q$ is greater than what
we assumed; or (2) they currently experience external perturbations
which maintain their orbital eccentricities against tidal
dissipation.
%
\section{Constraints on the Tidal $Q$ value of close-in  planets}
Now we use Eq.~\ref{tcirc} to place constraints on the tidal
$Q$ values of the planets. An {\it upper\/} limit on $Q$ is provided
for planets with {\it circular\/} orbits since the circularization
must have occurred within the lifetime of the systems ($\tau_{\rm
circ} < \tau_{\rm age}$). Our assumption here is that these close-in
planets formed through tidal circularization of initially eccentric
orbits. \cite{Nagasawa08ap} showed that about one third of multiple
planetary systems could form close-in planets through tidal
circularization following a large eccentricity gain through
planet--planet scattering or Kozai-type perturbations. Direct
observational evidence for initially large orbital
eccentricities comes from the absence of
planetary orbits within {\it twice\/} the Roche limit around the
star \citep {Faber05,Ford06}.

On the other hand, close-in, {\em eccentric\/} planets impose a {\em
lower\/} limit on $Q$ values, since $\tau_{\rm circ}\geq \tau_{\rm
age}$ is expected for these systems, provided that they are not
currently subject to any eccentricity excitation mechanism.

For $P_{*,{\rm rot}}\sim 3-70\,$d, {\it all\/} planets in
Table~\ref{tb1} take $\Omega_{*,{\rm rot}}/n<(\Omega_{*,{\rm
rot}}/n)_{\rm crit}$, and hence $F_*>0$ and $\tau_{{\rm
circ},0}>\tau_{\rm circ}$. For planets with zero
(non-zero) eccentricity, we require $\tau_{\rm circ}<\tau_{{\rm
circ},0}<\tau_{\rm age}$ ($\tau_{\rm age}<\tau_{\rm circ}<\tau_{{\rm
circ},0}$).  In other words, we assume that zero (non-zero)
eccentricity planets have been (have not been) circularized within
the lifetime of the system, independent of the rotation period of
the star. This gives the upper and lower limits for circular and
eccentric planets, respectively, as
\begin{eqnarray}
\Qp^{\prime}&<&
\frac{81}{2}n\left(\frac{M_*}{\Mp}\right)\left(\frac{\Rp}{a}\right)^5
\Fp \tau_{\rm age} \equiv Q_{\rm p, crit}^{\prime},
\end{eqnarray}
\begin{eqnarray}
\Qp^{\prime}&>& Q_{\rm p, crit}^{\prime}
\left(1-\frac{81}{2}\frac{n}{Q_*^{\prime}}\left(\frac{\Mp}{M_*}\right)
\left(\frac{R_*}{a}\right)^5 F_* \tau_{\rm age} \right)^{-1}.
\end{eqnarray}
The latter also gives the lower limit for the {\it
stellar\/} tidal $Q$ factor, since the denominator must be positive:
\begin{equation}
Q_*^{\prime}>\frac{81}{2}n\left(\frac{\Mp}{M_*}\right)\left(\frac{R_*}
{a}\right)^5 F_* \tau_{\rm age} \equiv Q_{*,{\rm min}}^{\prime}
\label{stellarQmin} \ .
\end{equation}
This corresponds to a minimum stellar $Q$ value of $Q_*^{\prime}\sim
3\times 10^4 - 4\times 10^7$, with a median value of
$0.4-1\times 10^6$ for $P_{*,{\rm rot}}=3-70\,$d,
which agree well with observations
\citep[e.g.][]{Carone07}.

\begin{figure}
\plotone{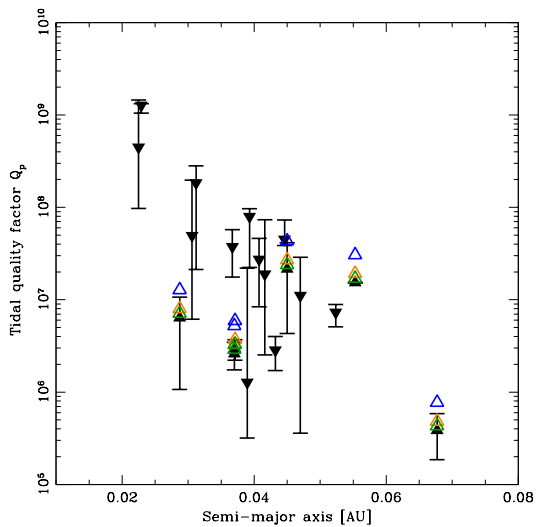} \caption[fig2]{Estimated tidal $Q$ factors for the
case of slowly rotating stars ($\Omega_*/n<(\Omega_*/n)_{\rm
crit}$). Upper/lower limits calculated from Eq.~\ref{tcirc0} are
shown in black down/up triangles for planets with zero/non-zero
eccentricities. Open triangles are the corresponding estimates from
Eq.~\ref{tcirc}, which approach black ones as we take 2, 5, and 10
times the minimum stellar tidal $Q$s (blue, orange, and green
triangles, respectively) obtained from Eq.~\ref{stellarQmin}.
\label{fig2}}
\end{figure}

Figure~\ref{fig2} shows the upper and lower limits for the planets'
$Q$ values as a function of semi-major axis for
circular and eccentric orbits, respectively.
Since the lower limits on $Q$ values for eccentric planets depend on
$Q_*^{\prime}$, we take three different cases of $Q_*^{\prime}=2$,
$5$, and $10\, Q_{*,{\rm min}}^{\prime}$ as examples. Note that,
with this definition of $Q_*^{\prime}$, $\Qp^{\prime}$ becomes
independent of the stellar spin rate. Since circularization times
are shorter for planets with smaller orbital radii, we tend to
overestimate the maximum $Q$ values at the shortest-period end. All
transiting planets appear within the range 
$10^5\lesssim \Qp^{\prime} \lesssim
10^9$. The figure also shows that the high eccentricities of some
planets (marked with upper triangles) can be explained by assuming
relatively large ($\Qp^{\prime}\gtrsim 10^6$) but reasonable
($\Qp^{\prime}\lesssim 10^9$) tidal $Q$ values.

Although these estimated $Q$ values are larger than those of
Jupiter or Neptune, they cannot be excluded. Recent theoretical
studies of the excitation and dissipation of dynamical tides within
rotating giant planets have shown that tidal $Q$ values fluctuate
strongly depending on the tidal forcing frequency, and the effective
$Q$'s could go up to $\sim10^9$ depending on the spin rate and internal
structure of the planet \citep[e.g.,
presence/absence of a core, radiative envelope, or a density jump,
see][]{Ogilvie04,Wu05b}. According to these recent models, it
appears possible that some planets maintain large eccentricities
simply because of their larger $Q$ values.
%
\section{Dynamical Perturbations}

Candidate perturbation mechanisms that could excite and maintain
planetary eccentricities include (1) tidal interaction with the
central star \citep{DobbsDixon04}, (2) quadrupole or higher-order
secular perturbation from an additional body, or (3) resonant
interaction with another planet. Here we discuss  the effects of
these competing mechanisms against tidal eccentricity damping.

Tidal dissipation inside the central star can increase the planet
eccentricity only when $\Omega_{*,{\rm rot}}>n$, or equivalently
$de/dt>0$  in Eq.~\ref{tcirc}. For a synchronized planet
($\Omega_{p,{\rm rot}}\sim n$) with a small eccentricity ($e^2\ll
1$), we obtain $\Omega_{*,{\rm
rot}}/n>18/11(1+7/18(Q_*^{\prime}/\Qp^{\prime})(M_*/\Mp)^2(\Rp/R_*)^5)$.
For a Jupiter-like planet around a main-sequence star, this yields
$\Omega_{*,{\rm rot}}>9.6n$ for $Q_*^{\prime}\sim \Qp^{\prime}$.
Since the rotation period for planet-hosting stars typically lies in
the range 3\,--\,70\,d, eccentricity excitation may occur for
planets only if their orbital periods are within 29\,--\,673\,d or
longer. For a $10 \MJ$ planet with radius $1 \RJ$, we have
$\Omega_{*,{\rm rot}}>1.7n$, which corresponds to an orbital period
greater than 5.1\,d. Therefore, this is unlikely to be responsible
for the eccentricity of observed planets within $a\sim
0.06$--0.18\,AU.

Another possibility is an undetected additional planet exciting the
eccentricity of the detected planet.  If there is a large mutual
inclination angle ($i \ga 40^\circ$) between the two planets,
Kozai-type perturbations can become important \citep{Kozai62}.  Such
highly non-coplanar orbits could result from
planet--planet scattering after dissipation of the gaseous disk.
\citet{Chatterjee07ap,Nagasawa08ap} have performed extensive
numerical scattering experiments and showed that the final
inclination of planets could be as high as $70^\circ$, with a median
of 10\,--\,$20^\circ$.  If $i \la 40^\circ$, octupole perturbations
may still moderately excite the eccentricity of the close-in planet.
The secular interaction timescale of a pair of planets with small
mutual inclination can be derived from the classical
Laplace-Lagrange theory \citep{Brouwer61,Murray99}.

For this secular perturbation from an additional planet to be
causing the large eccentricity of the close-in planet, it must occur
fast enough compared to other perturbations causing orbital
precession.  In particular, GR precession and tides are important
effects that would compete against the perturbation from the
additional body \footnote{Although stellar and planetary rotational
distortions cause additional precession \citep{Sterne39}, GR
precession dominates unless the stellar rotation is on the high
end.}. For detailed discussions see \cite{Holman97,Kiseleva98}.

\begin{figure}
\plotone{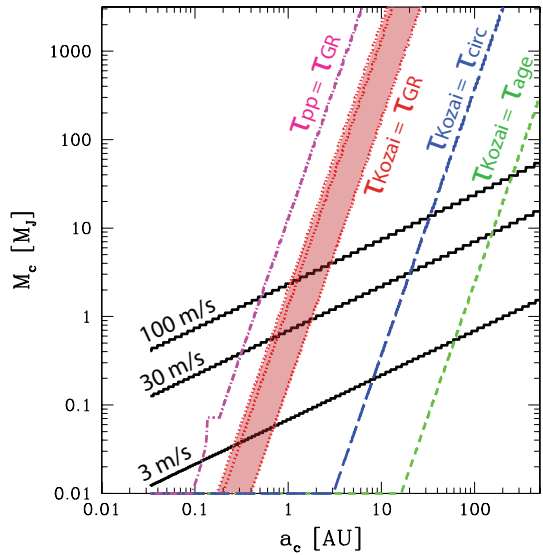}\caption[fig3]{Various secular eccentricity
excitation timescales for the planet GJ\,436\,b caused by a
hypothetical planetary (or stellar) companion GJ\,436\,c with mass
$\Mc$ and semi-major axis $\ac$. The solid black lines show the
predicted radial velocity amplitudes caused by the undetected
companion.  The dot-dashed line shows the threshold right of which
the secular interaction between planets with $i\lesssim40$ deg is
suppressed by GR precession, while the three red dotted lines are
the similar thresholds for the Kozai mechanism with the assumed
orbital eccentricity of the companion $\ec=$ 0.01, 0.5, and 0.9 from
left to right. We use $\tau_{\rm pp}$, $\tau_{\rm Kozai}$, and
$\tau_{\rm GR}$ as in \cite{Takeda08ap,Fabrycky07}. Thresholds for
$\tau_{\rm circ}$ and $\tau_{\rm age}$ are also shown for
comparison. \label{fig3}}
\end{figure}

Figure~\ref{fig3} illustrates the constraints on the mass and
orbital radius of the hypothetical outer planet in the GJ\,436
system. These are set by comparing the GR precession and secular
timescales. Similar results are seen for other systems with
eccentric close-in planets. Generally GR precession occurs faster
than any other perturbation mechanism. In order to induce Kozai
cycles in the inner planet (left of the dotted lines) while not
causing radial-velocity amplitudes above the detection limit of
$\sim5\,{\rm m} \,{\rm s}^{-1}$ (below black lines), the mass upper
limit of the hypothetical planet is $\sim 1 M_{\rm Neptune}$. For
near-coplanar systems even tighter constraints are placed on the
properties of the secondary planet (left of the dot-dashed line).

However, one caveat is that our diagram only rules out the
possibility of hypothetical bodies being {\it currently\/}
responsible for the high eccentricity of GJ\,436\,b. Also, while
Kozai-type perturbations are almost always suppressed by GR
precession, eccentricity excitation through secular {\it octupole\/}
perturbations may be occasionally {\em enhanced\/} by GR effects
\citep{Ford00,Adams06}. For the case of the GJ\,436 system, we have
numerically tested the effect of a hypothetical secondary planet~c
on the eccentricity evolution of the inner planet~b.  We have found
that, within the detectable radial-velocity limit $\la 5\,m/s$,
planet~c cannot excite the eccentricity of planet~b from 0.01 to the
observed 0.15, even if its eccentricity is as high as 0.5. This
result holds for other systems since they have even heavier planets.
Therefore, we can safely exclude the possibility that these planets
obtain their current high eccentricities through secular
perturbation from an undetected outer planet with $\lesssim M_{\rm
Neptune}$ if their orbits are initially near-circular.

Yet another possibility is a resonant perturbation from an
undetected planet.  Recently \citet{Ribas08} suggested that the
eccentricity of GJ\,436\,b might be caused by a mean-motion
resonance (MMR) with an unseen super-Earth, but there is little
observational support for this \citep[e.g.][]{Bean08ap}. Also, the
combined effects of GR precession and MMR are not fully understood
yet. In any case, it is unlikely that such resonances are
responsible for all the close-in eccentric planets, considering the
small fraction of extrasolar multiple planets in MMR.
%
%
\section{Summary}

In this letter, we have investigated the origins of close-in planets
on an eccentric orbit.  We place constraints on the tidal Q factor
of transiting planets by comparing the stellar age with the tidal
circularization time, and find that $10^5\lesssim \Qp^{\prime}
\lesssim 10^9$, which agrees well with current theoretical
estimates, can explain these eccentric planets. We also show that it
is difficult to explain the high eccentricities of these planets by
invoking a current interaction with an unseen second planet.  Our
results suggest that at least some of the close-in eccentric planets
may be simply in the process of getting circularized.

This work was supported by NSF Grant AST-0507727.
%
%


\begin{deluxetable}{lccccccc}
%
\tablecaption{Data are from {\tt http://exoplanet.eu/}. Ages are
computed using the stellar evolution database in \cite{Takeda07},
unless marked with $\ast$.
Median values of the derived posterior age probability distribution
functions are presented here, together with the 95\% credible
intervals in parenthesis. \label{tb1}}
\tablecolumns{8} \tablehead{ \colhead{Planet ID} & \colhead{$\Mp$
[$M_J$]} & \colhead{$\Rp$ [$R_J$]} & \colhead{$a$ [AU]} &
\colhead{e} & \colhead{$M_*$ [$M_{\odot}$]} & \colhead{$R_*$
[$R_{\odot}$]} & \colhead{Age [Gyr]} }
%
\startdata
OGLE-TR-56 b   &     1.29   &           1.3    & 0.0225     &
0    &    1.17     &      1.32  &  0.92 (0.20\,--3.00) \\
OGLE-TR-113 b   &     1.32   &      1.09 & 0.0229 & 0 & 0.78 & 0.77
&   13.28  (11.00\,--\,13.92) \\
GJ 436 b $\ast$  &    0.072   &          0.38    &    0.02872     &
0.15   &    0.452   &       0.464  &  6.00  (1.00\,--\,10.00) \\
OGLE-TR-132 b   &     1.14    &         1.18   &     0.0306
&       0   &     1.26    &  1.34  &  0.96 (0.12\,--\,3.84) \\
HD 189733 b   &     1.15   &        1.156  &      0.0312    &
0    &     0.8   &       0.753  &  8.96 (1.04\,--\,13.72) \\
TrES-2 $\ast$  &     1.98       &      1.22   &     0.0367   &
0   &     0.98     &         1  &  5.10 (2.40\,--\,7.80) \\
WASP-14 b $\ast$  &    7.725  & 1.259  &  0.037   &    0.095  &
1.319 &  1.297  &  0.75 (0.5\,--\,1) \\
WASP-10 b $\ast$  &    3.06 & 1.29 & 0.0371   &   0.057  & 0.71 & 0.783
& 0.8 (0.6\,--\,1) \\
HAT-P-3 b $\ast$  &     0.599     &        0.89 & 0.03894    &
0   &    0.936    &      0.824  &  0.40 (0.10\,--\,6.90) \\
TrES-1     &     0.61    &        1.081     &   0.0393   &     0
&      0.87    &       0.82  &  11.40 (3.20\,--\,13.84) \\
HAT-P-5 b $\ast$  &     1.06    &         1.26    &    0.04075     &
0    &    1.16    &      1.167  &  2.60 (0.80\,--\,4.40) \\
OGLE-TR-10 b   &     0.63   &          1.26  &      0.04162
&        0   &     1.18   &        1.16 &   1.20 (0.16\,--\,4.64) \\
HD 149026 b $\ast$   &    0.36    &         0.71    &    0.0432    &
0    &     1.3   &        1.45  &  2.00 (1.20\,--\,2.80) \\
HAT-P-4 b $\ast$ &      0.68     &        1.27     &   0.0446     &
0   &     1.26     &      1.59  &  4.20 (3.60\,--\,6.80) \\
HD 209458 b  &     0.69     &        1.32    &    0.045    &
0.07   &     1.01     &      1.12  &  2.40 (0.48\,--\,4.60) \\
OGLE-TR-111 b  &      0.53     &       1.067    &    0.047
&         0   &     0.82     &     0.831  &  5.17 (0.17\,--\,13.41) \\
HAT-P-6 b $\ast$  &    1.057    &         1.33    &    0.05235    &
0   &     1.29   &        1.46  &  2.30 (1.60\,--\,2.80) \\
HAT-P-1 b $\ast$  &     0.524    &         1.36   &     0.0553    &
0.067    &    1.133    &       1.115 & 3.60  \\
HAT-P-2 b $\ast$  &     8.64    &        0.952   &     0.0677 &   0.517
&    1.298    &      1.412  &  2.70 (1.30\,--\,4.10) \\
\enddata
\end{deluxetable}


\end{document}